\documentstyle[12pt]{article}
\begin{document}
\title{
\begin{flushright}
{\small SFU-HEP-10-94 \\November, 1994 }
\end{flushright}
\vspace{2cm}
Planckian Energy Scattering,\\ Colliding  Plane Gravitational
Waves and \\ Black Hole Creation
}

\author{ I.Ya. Aref'eva \thanks{Permanent address:  Steklov Mathematical
Institute, Vavilov st.42, GSP-1,117966, Moscow , Russia; E-mail :
arefeva@qft.mian.su}, K.S. Viswanathan
 \thanks{E-mail: kviswana@sfu.ca}
 and I.V.Volovich\thanks{Permanent address:  Steklov Mathematical
Institute, Vavilov st.42, GSP-1,117966, Moscow , Russia; E-mail:
volovich@mph.mian.su}\\
Department of Physics, Simon Fraser University\\
Burnaby, British Columbia, V5A 1S6, Canada}
\date {$~$}
\maketitle
\begin{abstract}

In a series of papers Amati, Ciafaloni and Veneziano and 't Hooft
conjectured that black holes occur in the collision of two light particles
at planckian energies. In this paper we discuss a possible scenario
for such a process by using the Chandrasekhar-Ferrari-Xanthopoulos duality
between the Kerr black hole solution and  colliding plane gravitational waves.
We clarify issues arising in the  definition of transition amplitude from
a quantum state containing only usual matter without black holes to a
state containing black holes.
Collision of two plane gravitational waves producing a
space-time region which is locally isometric to an interior of
black hole solution is considered. The phase of the transition
amplitude from plane waves to white and  black hole is calculated by using
the Fabbrichesi, Pettorino, Veneziano and Vilkovisky approach.
An alternative extension beyond the horizon in which the space-time again
splits into two separating gravitational waves is also discussed.
Such a process is interpreted as the scattering of plane gravitational
waves through creation of virtual black and white holes.
\end{abstract}

\newpage
\vspace{1cm}
\section{Introduction}
\setcounter{equation}{0}

In recent years there has been important progress in our
understanding of Planckian-energy scattering in quantum field
theory and in string theory \cite {1}-\cite{7}. Investigation of
such scattering could provide us with better understanding of
many fundamental issues such as creation of singularities and
black-holes, loss of information,and string modification of theory
of gravity.

Amati, Ciafaloni and Veneziano \cite{1} and 't Hooft \cite{2} made
a conjecture  that
black holes
will occur in the collision of two light particles at Planck energies
with small impact parameters. In this paper we discuss a possible
scenario for such a process by using colliding plane gravitational waves.
It was argued  \cite{1,2} that at extremely high energies
interactions due to gravitational waves will dominate all other quantum
field theoretic interactions. One can then visualize the following
mechanism of creation of singularities and/or black holes in the process of
collision of such high energy particles. Each of the two
ultra-relativistic particle generates a plane gravitational wave.
Then these plane gravitational waves collide and
produce a singularity or a black hole. It is difficult to perform
calculations for such a process in a realistic situation. We shall
discuss here an idealized picture.  Any gravitational wave far away from
sources can
be considered as a plane wave. We assume that plane
waves already have been produced by ultrarelativistic particles
and then consider analytically the process of black hole formation
under interaction of these plane waves.

One has two dimensionless parameters in a scattering process at Planck
energies\cite{1,2}. If $E$ is the energy in the center of mass frame, then
one defines the Planckian energy regime to be $GE^{2}/\hbar\geq 1$, where
$G$ is the Newton constant. This means that one can treat the process
semiclassically. The other dimensionless parameter is $GE/b$, where $b$
is the impact parameter. If one takes $GE/b\ll1$ , one can use the eikonal
approximation. The elastic scattering amplitude $A(s,t)$ in the eikonal
approximation was found in \cite{1,2}. Fabbrichesi, Pettorino,
Veneziano and Vilkovisky (FPVV) \cite{7}
gave the representation
\begin{equation}
A(s,t) \sim s \int d^{2}b~ e^{iqb} e^{iI_{cl}(s,b)}.
\end{equation}
Here $s$ and $t$ are the Mandelstam variables, $q^{2}=-t$.
$I_{cl}(s,b)$ was taken to be the value of the
boundary term for the gravitational
action calculated on the sum of
two Aichelburg-Sexl shock waves,
\begin{equation}
I_{cl}(s,b)=G s \log b^2.
\end{equation}
This corresponds to a linearization of
the problem and one cannot see in such
an approximation black hole formation.
It is tempting to suggest that one
can use the FPVV approach  \cite{7} and
the formula of type (1) even
for small impact parameters to calculate
the phase of the transition
amplitude from a state containing two
particles to a state containing a black hole. In such a case
$I_{cl}(s,b)$ could be the value
of the gravitational action for a
solution of Einstein equation
corresponding to an interior of the Kerr
black hole created in the collision of
plane gravitational waves. The
parameters $s$ and $b$ in this case
could be expressed in terms of the
mass and angular momentum  of the Kerr black hole.

Classical collision of plane gravitational waves has been the
subject of numerous investigations, see for example
\cite{SKP,CX,FI,Gr}, and it has a remarkably rich structure.
Here we are going to
use the Chandrasekhar-Ferrari-Xanthopoulos duality between
colliding plane
gravitational waves and the Kerr black hole solution.

The paper is organized as follows. Sect.2 contains a discussion
concerning quantum mechanical
description
of black hole creation in the scattering processes of particles.
In Sect.3 we discuss the boundary
term in the gravitational action in the context
of semiclassical approximation and present the boundary term  in the form
suitable to calculate non-linear
effects in quantum collision of gravitational
waves. Sect.4 contains a necessary information about classical solutions
describing colliding plane waves.
In Sect.5 we present the semiclassical transition amplitude for the process
of creation of black hole in the collision of two plane waves. We show also
that  the transition amplitude for elastic scattering of two plane waves
is equal to $1$ in the leading order of the semiclassical approximation.
This conclusion is based  on the investigation of the
behaviour of geodesics for the metric
corresponding to elastic scattering of two plane waves.
Its behaviour is described  in Appendix.

\section{Black Hole   Transition Amplitude }
 \setcounter{equation}{0}

We discuss here some issues concerning quantum mechanical
description
of black hole creation in the scattering processes of particles.
Let us note first that black holes  cannot be incorporated into the
theory if we consider quantum field theory in  Minkowski space-time.
In fact it is obvious
from Einstein
equation
\begin{equation}
        R_{\mu \nu }-\frac{1}{2}g_{\mu \nu}R=8\pi GT_{\mu \nu}
	\label{2.1}
\end{equation}
that if one has  matter, i.e. a nonvanishing $T_{\mu \nu}$
then one has a nontrivial gravitational field. This means that
one has to
start with a region of space-time which is flat only in some
approximation. Then, in the process of collision one gets a
strong gravitational field including perhaps black holes and/or
singularities.
Let us clarify the
meaning of the transition amplitude from a state
describing particles to a
state containing black holes,
$$<\mbox {Black holes}|\mbox{Particles in almost
 Minkowski space-time}>.$$

 By the analogy with the definition
 of transition amplitude in quantum mechanics
this transition amplitude
can be characterized  by values of metric and others fields in two
given moments of time (or by data on two given Cauchy surfaces,
say $\Sigma '$ and $\Sigma ''$).

Our starting point is the quantum-mechanical Feynman transition
amplitude between  definite configurations of the three-metric $h_{ij}'$
and field $\Phi '$ on an initial spacelike surface  $\Sigma ' $
and a configuration $h_{ij}''$ and  $\Phi ''$ on a final surface
 $\Sigma '' $. This is
 \begin{equation}
<h'',\phi '', \Sigma ''|h', \phi ' , \Sigma '>=
\int ~~ e^{\frac{i}{\hbar}S[g,\Phi]}
{}~{\cal D}\Phi {\cal D}g,                               \label {3}
\end{equation}
where the integral is over all four-geometries and field configurations
which match  given values on two spacelike surfaces, i.e.
 \begin{equation}
\Phi |_{\Sigma '}=\phi ',~
g |_{\Sigma '}=h '                                    \label {4}
\end{equation}
 \begin{equation}
\Phi |_{\Sigma ''}=\phi '',~
g |_{\Sigma ''}=h ''                                    \label {5}
\end{equation}
In this paper we shall consider the transition amplitude
in the semiclassical approximation and we don't
discuss here the introduction of ghosts and the
definition of the measure $Dg$.

We are interested in the process of black hole creation. Therefore we
specify the initial configuration $h' $ and $\phi ' $ on $\Sigma '$
as configuration of gravitational and matter fields in Minkowski
spacetime and we specify the final configuration $h''$ and $\phi ''$
on $\Sigma ''$
as describing black hole.
So, $\Sigma '$ {\it is a partial Cauchy surface with
 asymptotically simple
past in a strongly asymptotically predictable space-time and }
$\Sigma ''$ {\it is a partial Cauchy surface
 containing black hole(s), i.e.
$\Sigma ''-J^{-}({\cal T}^{+})$ is non empty}.
To explain these let us recall some necessary notions
from the theory of black holes \cite {HE,Wa}.

Black holes  are conventionally defined in
asymptotically flat space-times by the
existence of an event horizon $H$.
The horizon $H$ is the boundary $\dot {J}^{-}({\cal I}^{+})$
 of the causal past $J^{-}({\cal I}^{+})$
of future null infinity ${\cal I}^{+}$, i.e. it is the boundary of the
set of events
in space-time from which one can escape to infinity in the future
direction.

The {\it black hole region } $B$ is $$B=M-J^{-}({\cal I}^{+})$$
and {\it the event horizon}
$$H=\dot {J}^{-}({\cal T}^{+}).$$

Consider a space that is asymptotically flat in the sense of being
{\it weakly asymptotically simple and empty}, that is , near future and
past null infinities it has a conformal structure like that of Minkowski
space-time. One assumes that  space-time is {\it future asymptotically
predictable}, i.e. there is a surface ${\cal S}$ in spacetime that serves
as a Cauchy surface for a region
extending to future null infinity. This means that there are no "naked
singularities"
(a singularity that can be seen from infinity)
to the future of the surface ${\cal S}$.
This gives a formulation of Penrose's cosmic censorship
conjecture.

A space-time $(M, g_{\mu \nu})$
is {\it asymptotically simple} if there
exists a smooth manifold $\tilde {M}$
with metric $\tilde {g}_{\mu\nu}$,
boundary ${\cal T}$, and a scalar
function  $\Omega $ regular everywhere on
$\tilde {M}$ such that

(i) $~~\tilde {M}-{\cal I}$ is conformal
to $M$ with  $\tilde {g}_{\mu\nu}=
\Omega ^{2} g_{\mu\nu}$,

(ii) $~\Omega >0$ in $\tilde {M}-{\cal T}$
and     $\Omega =0$ on ${\cal I}$
with $\nabla _{\mu }\Omega \neq 0$ on ${\cal I}$,

(iii) Every null geodesic
on $\tilde {M}$ contains, if maximally extended,
two end points on  ${\cal I}$.

If $M$ satisfies the Einstein vacuum equations near ${\cal I}$
then ${\cal I}$ is null. ${\cal I}$
consists of two disjoint pieces ${\cal I}
^{+}$ (future null infinity)
and ${\cal I}^{-}$ (past null infinity)
each topologically $R$x$S^{2}$.

A space-time $M$ is {\it weakly asymptotically simple}
 if there exists
an asymptotically simple $M_{0}$ with corresponding $\tilde{M}_{0}$
such that for some open subset K of  $\tilde{M}_{0}$
including ${\cal I}$,
the region $M_{0}\cup K$ is isometric to an open
 subset of $M$.  This
allows $M$ to have more infinities than just ${\cal I}$.

The domain of dependence $D^{+}(\Sigma)$ ($D^{-}(\Sigma)$)
of a set $\Sigma$ is defined as the set of all points $p\in M$
such that every  past (future) inextendible non-spacelike
 curve through $p$
intersects $\Sigma$. A space like hypersurface which no non-spacelike
curve intersects more than once is called a partial Cauchy surface.
Define $D(\Sigma) =D^{+}(\Sigma)\cup D^{-}(\Sigma)$.
A partial Cauchy surface $\Sigma$ is said to be a global
 Cauchy surface
if $D(\Sigma)=M$.

Let $\Sigma $ be  a partial Cauchy surface in a weakly
 asymptotically simple
and empty space-time $(M,g)$. The space-time $(M,g)$ is (future) {
\it asymptotically
predictable from} $\Sigma$ if ${\cal I}^{+}$ is contained
in the closure
of $D^{+}(\Sigma)$ in $\tilde{M}_{0}$.
If, also, $J^{+}(\Sigma) \cap \bar{J^{-}}({\cal I}^{+},\bar{M})$
is contained in $D^{+}(\Sigma)$ then the space-time $(M,g)$
is called strongly   asymptotically predictable  from $\Sigma$.
In such a space there exist a family $\Sigma (\tau)$, $0<\tau<\infty$,
of spacelike surfaces homeomorphic to $\Sigma$ which cover
$D^{+}(\Sigma)-\Sigma$ and intersects ${\cal I}^{+}$.
 One could regard
them as surfaces of constant time.
A {\it black hole on the surface } $\Sigma (\tau)$ is
 a connected component of the set
$$
B(\tau)=\Sigma (\tau)- J^{-}({\cal I}^{+},\bar{M}).$$

One is interested primarily in
black holes which form from an initially
non-singular state. Such a state
can be described by using the partial
Cauchy surface $\Sigma$ which has an {\it asymptotically simple past},
i.e.
the causal past $J^{-}({\Sigma})$ is isometric to the region
 $J^{-}(\cal I)$ of some asymptotically
 simple and empty space-time with
 a Cauchy surface ${\cal I}$. Then $\Sigma$ has the topology $R^{3}$.

In the case considered one has space-time
$(M,g_{\mu\nu})$ which is weakly asymptotically simple and empty
and  strongly asymptotically
predictable  .

$\Sigma ' $ is a partial Cauchy surface with asymptotically simple past,
$\Sigma ' \sim R^{3}$.

 $\Sigma ''=\Sigma (\tau '') $ contains a black hole, i.e. $\Sigma '' -$
$J^{-}({\cal I} ^{+}, \bar{M})$ is nonempty.

In particular, if one has the
condition $\Sigma ' \cap \bar {J}^{-}({\cal I})$
is homeomorphic to $R^{3}$ (an open set with compact closure)
then $\Sigma ''$ also satisfies this condition.

Strictly speaking one cannot apply the standard definition of black holes
to the case of  plane gravitational waves and we need a
generalization.One defines a black hole in terms of the event horizon,
 $\dot {J}^{-}({\cal I} ^{+})$.
However this definition depends on
the whole future behaviour of the metric.
There is  a different sort of horizon
which depends only on the properties
of space-time on the surface  $\Sigma (\tau) $ \cite {HE}.
Any point in the black hole region bounded by $r=2m$ in the Kruskal
diagram represents a {\it trapped surface} (which is a two-dimensional
sphere in space-time) in that both the outgoing and ingoing families of
null geodesics emitted from this point converge and hence no light ray
comes out of this region. A generalization of the definition of black
holes in terms of trapped horizon has been considered in \cite
{Tipler,Hay}.
A generalization of the standard definition of black holes to the case of
nonvanishing cosmological constant was considered in \cite{GibHaw}.

We discussed the transition amplitude (propagator) between definite {\it
configurations} of fields,
$<h'',\phi '', \Sigma ''|h', \phi ' , \Sigma '>$.
The transition amplitude from {\it a state} described by the wave
function $\Psi ^{in}[h', \phi ']$ to a state $\Psi ^{out}[h'', \phi '']$
reads
\begin{equation}
<\Psi ^{out}|\Psi ^{in}>=
\label {7}
\end{equation}
$$\int \bar{\Psi }^{out}[h'', \phi '']
<h'',\phi '', \Sigma ''|h' ,\phi ' , \Sigma '>\Psi ^{in}[h' ,\phi ']
{\cal D}h'{\cal D}\phi '{\cal D}h''{\cal D}\phi ''.$$
One can take for example the state $\Psi ^{in}=
\Psi ^{in}_{p_{1}p _{2}}[h', \phi ']$
as a Gaussian distribution describing a state of particles with momenta
$p_{1}$ and $ p_{2}$ and take $\Psi ^{out}$ as a wave function describing a
state of black hole. Recently Barvinski, Frolov and Zelnikov
have suggested an expression for the wave function of the ground state of a
black hole \cite {Frolov}.

\section{Boundary term in Gravitational Action
and Semiclassical  Expansion }
\setcounter{equation}{0}

In this section we discuss an approximation scheme forcalculating
the transition amplitude following to the FPVV approach \cite{7}.
The gravitational action with the boundary term has the form \cite{York}
\begin{equation}
S[g]=\frac{1}{16\pi G}\int_{V} R\sqrt{-g}d^4x -
\frac{1}{8\pi G}\int_{\partial V} K\sqrt{h}d^3x \label {3.1}
\end{equation}
Here $V$ is a domain in space-time
with the space-like  boundary   $\partial V$,
$h$ is the first fundamental form and $K$ is the trace of the
second fundamental form of $\partial V$.

The case of null surfaces was considered in \cite {Isr}.
We shall
write a representation of the
boundary term suitable for quantum consideration.
The action is
 \begin{equation}
S[g]=-\frac{1}{16\pi G}\left(\int_{V} d^{4}x \sqrt{-g}~ R (g)+
\int _{V}d^{4}x \sqrt{-g}~\nabla _{\mu}f^{\mu}(g)\right),
                                                      \label {3.2}
\end{equation}
where
 \begin{equation}
                                                       \label {3.3}
f^{\mu}(g)=g^{\alpha\beta}g^{\mu \nu}
\partial_{\nu} g_{\alpha\beta}-
g^{\mu\alpha}g^{\beta \nu}\partial_{\nu}g_{\alpha\beta}.
\end{equation}
Supposing that the boundary is described by equation
 \begin{equation}
 \sigma (x) =0
                \label {3.10}
\end{equation}
one gets the action in the form
 \begin{equation}
                \label {3.11'}
S[ g_{cl}]=-\frac{1}{16\pi G}(\int _{V}d^{4}x \sqrt{-g}~ R (g)+
\int _{V}d^{4}x \sqrt{-g_{cl}}~\delta (\sigma (x))~
f^{\mu}\nabla_{\mu}\sigma ).
\end{equation}

The linearization of the action (\ref {3.2}) leads to the action
in the FPVV form  \cite {7}.

Let us show that the presence of the boundary term in the action
(\ref {3.2}) is necessary for the selfconsistency of the semiclassical
expansion. To perform semiclassical
expansion one expands the metric $g$ around a
classical solution $g_{cl}$ of the Einstein
equation so that $g=g_{cl}+\delta g$,
 \begin{equation}
S[g]=S[g_{cl}+\delta g]=S[g_{cl}]+S'[g_{cl}]\delta g
+\frac{1}{2}S''[g_{cl}](\delta g)^{2}+...,
\label {3.3'}
\end{equation}
where $g_{cl}$ matches the given Cauchy data on surfaces
$\Sigma '$ and $\Sigma ''$, i.e.
 \begin{equation}
g  _{cl}~|_{\Sigma '}=h ' ,~~\label {3.4}
g  _{cl}~|_{\Sigma ''}=h ''
\end{equation}
In this case $\delta g|_{\Sigma '}=\delta g|_{\Sigma ''}=0$, but we
cannot guarantee that $\nabla\delta g|_{\Sigma '}=\nabla\delta g|_{\Sigma
''}=0$.
To ensure that terms linear on $\nabla\delta g$
drop out from expression (\ref{3.3'}) (otherwise we cannot perform
semiclassical expansion) one has to integrate by parts.  One has
\begin{equation}
S[ g_{cl}+\delta g]=-\frac{1}{16\pi G}\left(\int d^{4}x
\sqrt{-g_{cl}}~R
(g_{cl})+
\int d^{4}x \sqrt{-g_{cl}}~\nabla _{\mu}f^{\mu}(g_{cl})) + \right.
 \label {3.8}
 \end{equation}
$$+\left. \int d^{4}x \sqrt{-g_{cl}}~(g^{\mu \nu}
\nabla ^{2}\delta g_{\mu \nu}
-\nabla ^{\mu}\nabla ^{\nu}\delta g_{\mu \nu})+
\nabla _{\mu}( g^{\alpha\beta}\partial^{\mu}\delta g_{\alpha\beta }
 g^{\mu\alpha}-\partial ^{\beta}\delta g_{\alpha\beta })\right)
$$
$$+~ second ~~order~~terms.
$$
The linear terms coming from  the Hilbert-Einstein action
 can be put in the form
 \begin{equation}
                          \label {3.9}
g^{\mu \nu} g^{\alpha\beta}\nabla _{\alpha}
\nabla _{\beta}\delta g_{\mu \nu}
-\nabla ^{\mu}\nabla ^{\nu}\delta g_{\mu \nu}=
\nabla _{\mu}( g^{\alpha\beta}\nabla^{\mu}\delta g_{\alpha\beta }
 -g^{\mu\alpha}\partial^{\beta}\delta g_{\alpha\beta })
\end{equation}
Notice that on the RHS of the last relation covariant derivatives
can be replaced by partial  derivatives because
on the boundary $\delta g =0$.
Therefore one finds that terms linear in
$\delta \partial g $ coming from Hilbert-Einstein action
cancel  similar terms coming from
full divergence and the action (\ref {3}) admits
the expansion (\ref {3.3'}).

Taking into account that the value of the Hilbert-Einstein action
 on the classical
solution is equal to zero one finds that the full action for a solution of
Einstein equation is
reduced to the second term in (\ref{3.8}) which
can be reduced to  boundary term.

The transition amplitude (\ref{3}) in semiclassical approximation is
 \begin{equation}
<h'',\Sigma ''|h' ,\Sigma '>={\cal Z} \exp {\frac{i}{\hbar }S_{cl}}
\label {3.12}
\end{equation}
where
 \begin{equation}
                \label {3.11}
S[ g_{cl}]=-\frac{1}{16\pi G}\int d^{4}x \sqrt{-g_{cl}}~\delta (\sigma (x))~
f^{\mu}\nabla_{\mu}\sigma ,
\end{equation}
 $f$ is given by equation (\ref{3.3}) and
$$
{\cal Z}=(\frac{\pi}{\det S''(g_{cl})})^{1/2}
$$
We assume here that there is only one solution
of classical equation of motion with given boundary conditions.

\section{Colliding Plane Gravitational
Waves }
\setcounter{equation}{0}

\subsection{Szekeres line element}
In this section we present a solution of Einstein
equations describing colliding plane gravitational waves.
There exists a well known class of plane-fronted gravitational waves
with the metric
\begin{equation}
        ds^{2}=2dudv +h(u,X,Y)du^{2}-dX^{2}-dY^{2}
	\label{o}
\end{equation}
where $u$ and $v$ are null coordinates .
In particular the gravitational field
of a particle moving with the speed of
light is given by the Axelburg-Sexl solution. The metric has the form
\begin{equation}
ds^{2}=2dudv + E\log(X^2+Y^2)\delta(u)du^{2}-dX^{2}-dY^{2}
        \label{oo}
\end{equation}
and describes a shock wave. It is difficult to find a solution which
describes two sources.  An approximate solution of
equation (\ref{2.1}) for two particles as the sum of solutions each of
which describes one particle was considered by
FPVV\cite {7}. This
approximation describes well the scattering amplitude for large impact
parameter, but does not describe non-linear interaction of shock waves
which is dominate in the region of small impact parameter. To analyze
non-linear effects we will, instead of dealing with shock wave,  take
a  simple solution of Einstein equation, namely we will
take  plane gravitational
waves. In some respects
one can consider plane wave as an approximation to  more complicated
gravitational waves, in particular shock waves.
This solution in some sense can be interpreted as an approximation for a
solution of Einstein equation in the presence of two
moving particles.
Collision of two ultrarelativistic black holes was considered
by D'Eath \cite{DE}.
If our particles are gravitons then there are not
sources corresponding to matter fields in Einstein equations.
Note also that plane gravitational waves are produced
by domain walls, see \cite {4}.

A particular class of
{\it plane waves} is defined to be
plane-fronted waves in which the field components are the same at every
point of the wave surface. This condition requires that $h(u,X,Y)$
is a function with a quadratic dependence on $X$ and $Y$.
One can then remove the dependence of h on $X$\ and $Y$ altogether by
a coordinate change.
Solutions of Einstein equations describing {\it collisions } of
plane gravitational waves
were first obtained by Szekeres  and Khan and Penrose
\cite{SKP}.  Chandrasekhar, Ferrari and Xanthopoulos
\cite {CX,FI} have developed a powerful method for obtaining such solutions
by using a remarkable analogy ("duality") with stationary axisymmetric
case  which
 can be reduced to the investigation of the Ernst equation. For
a review see the Griffiths book \cite {Gr}.

We will use the coordinates $(u,v,x,y)$.
We assume that throughout  space-time there exists a pair of commuting
space-like Killing vectors $\xi _{1}=\partial _{x}$,  $\xi _{2}=\partial
_{y}$ . The Szekeres line element has the form
\begin{equation}
ds^{2}=2e^{-N}dudv -e^{-U}(e^{V}\cosh Wdx^{2}-
-2\sinh W dxdy+e^{V}\cosh Wdy^{2}),
	\label{4.3}
\end{equation}
Here $N,U,V$ and $W$ are functions of $u$ and $v$ only.

We illustrate in Fig.1 the two-dimensional geometry of plane waves.
Space-time is divided into four regions. The region {\bf I}
is the flat background before arriving plane waves.
The null hypersurfaces $u=0,$
and $v=0$ are the past wave fronts of the incoming plane waves 1 and 2.
The metric in
region {\bf I} is Minkowski. Regions {\bf II} and {\bf III}
represent incoming plane waves which interact
in  region {\bf IV}.  Colliding plane gravitational waves can
produce singularities or Cauchy horizons in the interaction region
\cite{MT,CX,Yu1,KH,Hay}. A solution is undetermined to the future
across a Cauchy horizon \cite{CX,Yu2,Hay,HEr}. We shall discuss
two simplest extensions.

In particular one can get an interiour of the Schwarzschild solution
in the interactiong region {\bf IV}.There are two types of colliding
plane waves solutions corresponding to the Schwarzschild metric. The
first one creates the interior of the blach hole with the usual
curvature singularity. In this case incoming plane waves have curvature
singularities already before collision. In the context of the Planck
energy scattering it seems more natural don't have curvature singularities
already for free plane gravitational waves. Therefore we mainly will be
discussing another type of solutions when one getsin the interaction region
the interior of the Schwarzschild white hole. The maximal analytic extension
of this solution across its Killing-Couchy horizon leads to creation of a
covering space of the Schwarzschild black hole out of collision between
two plane gravitational waves. An alternative interpretation of this
solution is the creation of the usual Schwarzschild black hole
out of the collizion between two plane gravitational waves propagatiing in
a cylindrical universe. There exist also a time-reversed extension
\cite {Hay} including covering space of the Schwarzschild exterior
and part of black hole, and giving two receding plane waves with flat
space between. We will interprete this as scattering of plane waves
on the virtual black hole.

\subsection{Ernst's equation for colliding  plane gravitational
waves}
Vacuum Einstein equations
\begin{equation}
        R_{\mu \nu}-\frac{1}{2} R g_{\mu\nu}=0
	\label{4.4}
\end{equation}
for the metric (\ref{4.3}) have the form
\begin{eqnarray}
	U_{uv} & = & U_{u}U_{v}
	\label{4.5} \\
	2	U_{vv} & = &
	U_{v}^{2}+W_{v}^{2}+ V_{v}^{2}\cosh ^{2}W - 2 U_{v}N_{v}
	\label{4.6} \\
  	2U_{uu}& =&U_{u}^{2}+W_{u}^{2}+ V_{u}^{2}\cosh ^{2}W - 2 U_{u}N_{u}
	\label{4.7} \\
        2U_{uv}& =&     U_{u}V_{v}+
        U_{v}V_{u} -2 (V_{u}W_{v}+ V_{v}W_{u})\tanh W
	\label{4.8} \\
	2N_{uv}& =&	-U_{u}V_{v}+W_{u}W_{u}+V_{u}V_{v}\cosh ^{2} W
	\label{4.9} \\
        2W_{uv}& =&     U_{u}W_{v}+U_{v}W_{u}+2V_{u}V_{v}\sinh W\cosh  W
        \label{4.10}
\end{eqnarray}
One can reduce the system of equations (\ref{4.5})-(\ref{4.10})
to a solution of the Ernst equation for the complex valued function
\begin{equation}
Z=\chi +i\lambda
	\label{4.12}
\end{equation}
where
\begin{equation}
\chi =e^{-V}\sinh W , ~~  \lambda = e^{-V}\tanh W       .
	\label{4.11}
\end{equation}
The line element becomes (\ref {4.3})
\begin {equation} 
                                                          \label {4.11'}
ds^{2}= 4e^{-N}dudv-e^{-U}[\chi dy^{2}+\frac{1}{\chi }
(dx-\lambda  dy)^{2}],
\end   {equation} 

First note that equation (\ref{4.5}) can be integrated to give
\begin{equation}
 e^{-U} =f(u)+g(v),
        \label{4.12'}
\end{equation}
where $f(u)$ and $g(v)$ are arbitrary functions. We fix a gauge by using
$f=f(u)$ and $g=g(v)$ as new coordinates instead of $u$ and $v$ and then
we change variables from $f$
and $g$ to $\mu $ and $\eta$ such that
\begin{equation}
f+g= (1-\mu ^{2})^{1/2} (1-\eta ^{2})^{1/2},~~f-g=-\mu\eta .
	\label{4.13}
\end{equation}
One has the Ernst equation
\begin{equation}
(Z+\bar{Z})[((1-\mu ^{2}Z_{, \mu})_{, \mu}-((1-\eta
^{2})Z_{,\eta})_{,\eta}]=2[(1-\mu ^{2}Z_{, \mu}^{2}-(1-\eta
^{2})Z_{,\eta}^{2}].
	\label{4.14}
\end{equation}
 If a solution $Z$  of the Ernst equation
(\ref{4.14})  is found then the metric functions $V$ and
$W$ are given by
\begin{equation}
e^{V}=(Z \bar {Z})^{-1},
        ~~\sinh W = -i \frac{Z-\bar {Z}}{Z+ \bar {Z}}.
                \label{4.16}
\end{equation}
$U$ is given by
$e^{U}=(1-\mu ^{2})^{1/2}(1-\eta ^{2})^{1/2}
 $
 and  $N$ is defined by the relation
$e^{-N}  = f' g'(f+g)^{-1/2}e^{-B}$.
To find  N one has to integrate the following equations
\begin{equation}
        B_{,f}  = -2 (f+g)(1-E \bar {E})^{-2}E_{,f}\bar {E}_{,f},
        ~~B_{,g} =  -2 (f+g)(1-E \bar {E})^{-2}E_{,g}\bar {E}_{,g},
                        \label{4.21}
\end{equation}
where $E$ is the Ernst function related with $Z$, $E=(Z-1)(Z+1)^{-1}$.

\subsection{ Chandrasekhar-Ferrari-Xanthopoulos duality}
The Ernst equation (\ref {4.14}) is invariant under the Neugebauer-Kramer
involution. If $Z_{0}=\chi +i\lambda $ is a solution of
(\ref {4.14}) then
\begin{equation}
Z=\Phi +i\Psi
	\label{43.1}
        \end{equation}
is also a solution of (\ref {4.14}), where $\Phi $ and $\Psi$ are given by
equations
\begin{equation}
\Psi = (1-\mu ^{2})^{1/2}(1-\eta ^{2})^{1/2}\chi ^{-1}.
	\label{43.2}
\end{equation}
\begin{equation}
\Phi _{,\eta} =\frac{1-\mu ^{2}}{\chi ^{2}}\lambda _{,\mu},
\Phi _{,\mu} =\frac{1-\eta ^{2}}{\chi ^{2}}\lambda _{,\eta}        \label{43.3}
\end{equation}
Notice here that for stationary axisymmetric space-time one has the same
Ernst equation (\ref {4.14}), whose solution
$Z=\Psi + i\Phi$ is considered as a potential for the metric functions $
\chi $ and $\lambda$ that are obtained from (\ref {43.3}). Take
	\begin{equation}
        E= p \mu  +iq \eta .
		\label{43.5}
        \end{equation}
This is a solution of Ernst's equation if
  \begin{equation}
	p^{2}+q^{2}=1
		\label{43.6}
	\end{equation}
If one considers (\ref {43.5}) as containing the metric functions, $E=
\chi +i\lambda$, it leads to the Nutki-Halil solution describing colliding
gravitational waves with non-aligned polarization.
However, when the same function (\ref {43.5}) is considered as a
potential, it then leads to the Kerr solution. Therefore a region of the
Kerr space-time can be also considered as a solution describing plane
waves. This remarkable observation we call  the
Chandrasekhar-Ferrari-Xanthopoulos
duality. The Ernst potential (\ref {43.5}) gives
\begin{equation}
Z=\Psi + i\Phi =\frac{1+p\mu +iq \eta}{1-p\mu -iq \eta}
	\label{43.7}
\end{equation}
and one gets the following solution of equations (\ref {43.2})
and (\ref {43.3})
\begin{equation}
        \chi   =  (1-\mu ^{2})^{1/2}(1-\eta ^{2})^{1/2}
        \frac{X}{Y}
        \label{43.8}
\end{equation}
\begin{equation}
\lambda   =  \frac{2q}{p}[\frac{1}{1+p }-
  \frac{(1-\eta ^{2})(1-p\mu)}{1-p ^{2}\mu ^{?}-q^{2}\eta ^{2}} ].
        \label{43.8'}
\end{equation}
Here
\begin{equation}
X= (1-p\mu) ^{2}+q^{2}\eta ^{2},~~Y= 1-p ^{2}\mu ^{2}-q^{2}\eta ^{2}
	\label{43.10}
\end{equation}

For the Ernest potential the metric  has the form
\begin{equation}
ds^{2}= X(
\frac{d \eta ^{2}}{1-\eta ^{2}}-\frac{d \mu ^{2}}{1-\mu ^{2}})-
(1-\mu ^{2})^{1/2}(1-\eta ^{2})^{1/2}[\chi dy^{2}+\frac{1}{\chi}
(dx-\lambda dy)^{2}].
        \label{43.9}
\end{equation}
 Now let us take
\begin{equation}
p=-\frac{(m^{2}-a^{2})^{1/2}}{m}, ~~ q= \frac{a}{m}, ~ m\geq a
	\label{43.11}
\end{equation}
 and introduce the new coordinates
 $(t,r,\theta , \phi )$ instead of $(\mu, \eta ,x,y)$ by putting
 \begin{equation}
\mu =\frac{r-m}{(m^{2}-a^{2})^{1/2}}, ~~ \eta =\cos \theta
 	\label{43.12}
 \end{equation}
 \begin{equation}
t= -\sqrt{2} m(x-\frac{2q}{p(1+p)}),~~ \phi =\frac{\sqrt{2} m}
{(m^{2}-a^{2})^{1/2}}y
        \label{4.13'}
\end{equation}
Then the metric (\ref {43.9}) will take the form of the Kerr solution
\begin{equation}
2m^{2}ds^{2}= (1-\frac{2mr}{\rho ^{2}})dt ^{2}-\frac{4amr}{\rho ^{2}}
\sin ^{2}\theta dt d\phi -
(r^{2}+ a ^{2}-\frac{2a^{2}mr}{\rho ^{2}})\sin ^{2}\theta  d\phi ^{2}
-\rho ^{2}(\frac{1}{\Delta} dr^{2} + d \theta ^{2})	,
	\label{43.14}
\end{equation}
where
\begin{equation}
\rho ^{2}= r^{2}+ a ^{2} \cos ^{2 }\theta, ~~\Delta= r^{2}-2mr+ a ^{2}
	\label{43.15}
\end{equation}
The coordinates must satisfy the inequality $|\eta|<\mu\leq 1$. This
implies that $-(m^{2}-a^{2})\sin ^{2}\theta <\Delta \leq 0 $, which means
the region of the Kerr spacetime that is inside the ergosphere.

To describe
the colliding plane
gravitational waves producing
the interior of the ergosphere in  the Kerr spacetime it is  convenient to
rewrite the metric (\ref {43.9}) in terms of the $(u',v')$ coordinates
related with $(\eta ,\mu )$ by the following relations
\begin {equation} 
                                                          \label {43.16}
\eta =\sin (u'-v'),~~\mu =\sin (u'+v'),
\end   {equation} 
These $(u',v')$ are related with $(u,v)$ in (\ref {4.3}) by $u=\sin u'$,
$v=\sin v'$. For simplicity of notations we will omit $'$ in (\ref {43.16}).
We have
\begin{equation}
ds^{2}= 4X(u,v)dudv-\Omega (u,v)[\chi (u,v)dy^{2}+\frac{1}{\chi (u,v)}
(dx-\lambda (u,v) dy)^{2}],
        \label{43.17}
\end{equation}
where
\begin {equation} 
                                                          \label {43.18}
X(u,v)=(1-p\sin (u+v))^{2}+q^{2}\sin ^{2}(u-v),~~\Omega (u,v)=
\cos(u+v)\cos(u-v),
\end   {equation} 
\begin {equation} 
                                                          \label {43.19}
\lambda (u,v)=\frac{2q}{1+p}\frac{1-\sin (u+v) }{Y(u,v)}((p+1)\sin ^{2}(u-v)+
p\sin (u+v)-1)
\end   {equation} 

\begin {equation} 
                                                          \label {43.20}
Y(u,v)=1-p^{2}\sin ^{2}(u+v)-q^{2}\sin ^{2}(u-v).
\end   {equation} 
In (\ref {43.17})
\begin{equation}
0<u<\pi/2,~~0<v<\pi/2,~~v+u<\pi/2
                                                          \label{43.21}
\end{equation}
To extend the metric (\ref {43.17}) outside of region (\ref {43.21})
one uses the Penrose-Khan trick and substitutes in (\ref {43.17})
\begin {equation} 
                                                          \label {43.22}
u\to u\theta (u),~~v\to v\theta (v),
\end   {equation} 
\begin{equation}
\theta (x)=\left\{
\begin{array}{cc}
	1, ~& x>0  \\
	0, ~& x<o
\end{array}
\right.
	\label{6.3}
\end{equation}
Fig.1 illustrates this metric.
The region {\bf I} ($u<0,~~v<0$) is Minkowskian.
Regions {\bf II} and {\bf III} contain the approaching plane
waves with the following metrics
\begin {equation} 
                                                          \label {43.23}
(ds^{II})^{2}= 4X(u)dudv-\Omega (u)[\chi (u) dy^{2}+\frac{1}{\chi (u)}
(dx-\lambda (u)dy)^{2}],
\end   {equation} 
\begin {equation} 
                                                          \label {43.24}
(ds^{III})^{2}= 4X(v)dudv-\Omega (v)[\chi (v) dy^{2}+\frac{1}{\chi (v)}
(dx-\lambda (v)dy)^{2}],
\end   {equation} 
where
\begin {equation} 
                                                          \label {43.25}
X(u)=(1-2p\sin u)+\sin ^{2}u,~~\Omega (u) =\cos^{2}u,
\end   {equation} 
\begin {equation} 
                                                          \label {43.26}
\lambda (u)=2q(\sin u -\frac{1}{1+p})
\end   {equation} 
In (\ref {43.23})
\begin{equation}
u<\pi/2,~~v<0,
                                                          \label{43.27}
\end{equation}
and in (\ref {43.24})
\begin{equation}
u<0,~~v<\pi/2,
                                                          \label{43.28}
\end{equation}
The region {\bf IV} is the interaction region with metric (\ref {43.17}).

\subsection{Plane Gravitational
Waves  for Zero Impact parameter}
In this subsection we discuss a particular example of two colliding plane
waves, namely $p=-1$, \cite {Hay}.
The choice is motivated by the fact that the colliding plane
gravitational waves produce in the interaction region a space-time that
is isometric to an interior of the Schwarzchild solution.
The metric is given by
$$ds^{2}=  4m^{2}[1+\sin (u\theta (u))+v\theta (v)]  dudv
        $$
\begin{equation}
                                                                \label{6.1}
	 -[1-\sin (u\theta (u))+v\theta (v)][1+
        \sin (u\theta (u))+v\theta (v)]^{-1} dx^{2}
 \end{equation}
$$         -[1+\sin (u\theta (u))+v\theta (v)]^{2}
	\cos ^{2}(u\theta (u))-v\theta (v))dy^{2},
$$
where
$u<\pi/2,~~v<\pi/2,~~v+u<\pi/2 .$

  Fig.1 illustrates  this solution of the vacuum Einstein equations.
The background
region {\bf I} describes a region of space-time before the arrival of
gravitational waves and it is Minkowskian.
Two planes  waves propagate
from  opposite directions along the z-axis.
Regions {\bf II} and {\bf III} contain the approaching plane
waves. In the region {\bf IV}
the metric (\ref{6.1})
is isomorphic to the Schwarzchild metric.
To see this one can make the following
change of variables from "plane waves" coordinates to Schwarzchild
coordinates,
\begin{equation}
(u,v,x,y) \to (t,r, \theta, \phi)
	\label{6.4}
\end{equation}
defined by,
\begin{equation}
r=m[1+\sin (u+v)],~~t=x,~~\theta= \pi/2 +u-v, ~~\phi =y/m ,
	\label{6.5}
\end{equation}
or to Kruskal coordinates $\tau, \zeta , \theta, \phi$
\begin{equation}
\tau =-a(r)\cosh t/4m,~~\zeta =- a(r)\sinh  t/4m,
	\label{6.6}
\end{equation}
where
\begin{equation}
a(r)=(1-r/2m)^{1/2}e^{r/4m}.
	\label{6.7}
\end{equation}
Then one gets
$$
ds^{2}=\frac{32m^{3}}{r}e^{-r/2m}
	 (d\tau ^{2} -d\zeta ^{2})
	 -r^{2}(d\theta ^{2} +\sin ^{2} \theta d\phi ^{2})
$$
 The section of the region {\bf IV} bounded
by  $x=0,~ y=0$ corresponds to segment in the Kruskal diagram
and the section of the region {\bf IV}
by the plane $x=x_{0},~ y_{0}=0$ corresponds to the hatched region
in the Kruskal diagram Fig.5. The lines corresponding to $r=2m$ (horizon)
apart from the point $(\tau =0,~\zeta =0)$ correspond to the infinite
value of $x$-plane wave coordinate.

The above metric in the $(u,v)$ plane can be extended beyond the event
horizon
$u+v=\pi/2$ in one of two ways.

The first possibility, shown in Fig.6 consists in reflecting along the
line $u+v=\pi/2$.

The second one involves in gluing  to the horizon the whole  upper-half
part of the Kruskal diagram.

Both extensions are solutions of Einstein equations. From the general
discussion in  Section 3 there is a-priori non-zero probability
to get a finite state corresponding to a black hole or two
outgoing plane waves. In the next section we will calculate the
probabilities for these processes
in the semi-classical approximation. As was mentioned in the Sections 2
and 3
to define correctly the finite Cauchy surface we have to know how the null
geodesics look like in space-time with the metric (\ref{6.1})
extended as shown in Fig.6.

 For the usual black hole solution, i.e. the
regions  {\bf IV } and ${\bf IV}^{'}$ are extended up to the boundary
$r=0$ and there are two asymptotically free regions, namely {\bf A}
and ${\bf A}^{'}$ in fig.5. In this case some geodesics cross the lines
$r=\pm 2m$ and may not return to region ${\bf VI}^{'}$.
So, it is important to know if it is possible for geodesics  starting from
Minkowski
region {\bf I} to end up in the region {\bf A} or ${\bf A}^{'}$.
A proof of this fact is given in Appendix.

\section{Semiclassical Transition Amplitude}
\setcounter{equation}{0}

In this section we  study the transition amplitude
\begin {equation} 
                                                          \label {5.1}
<BH|2pw>
\end   {equation} 
from a state $|2pw>$ of two plane gravitational waves to
a state $<BH|$ containing black-hole, and the transition amplitude
\begin {equation} 
                                                          \label {5.2}
<2pw|2pw>
\end   {equation} 
from two plane gravitational waves back  to two plane gravitational waves.
We can consider these two amplitudes as amplitudes of two independent
channels.

To find these transitions amplitudes in the semiclassical approximation
according to (\ref{3.12})
we have to evaluete the boundary term on the classical solution interpolated
 between
two plane gravitational waves in the initial state and
black hole in the final state, and on the classical solution
interpolated between
two plane gravitational waves in the initial state and
two plane gravitational waves in the final state.

Let us start from consideration  the transition
amplitude
\begin {equation} 
                                                          \label {5.2'}
<2pw, WH|2pw>
\end   {equation} 
from two plane gravitational waves  to two plane gravitational waves and
white hole. The corresponding classical solution is shown on fig.3.
In this case the initial Cauchy surface $\Sigma '$ crosses regions
{\bf I},{\bf II} and {\bf III} and the final surface crosses regions
{\bf II},{\bf IV} and {\bf III} (Fig.7). Suppose that the equation
(\ref {3.10})  for the  surfaces $\Sigma '$ and $\Sigma ''$
has the form
\begin{equation}
\Sigma ':  \sigma = u-v-w'_{0} =0, ~w'_{0}\leq 0
                                                         \label{7.1}
\end{equation}
 \begin{equation}
\Sigma '': \sigma = u-v-w''_{0}=0, ~0<w''_{0}\leq \pi /2
                                                         \label{7.2}
\end{equation}
In this case the other parts of the boundary defining the boundary term
are given by equations
  \begin{equation}
 B_{2}: ~u-\pi/2 =0,  ~w'_{0}-\pi /2 \leq v\leq  w''_{0}-\pi /2;~~
 B_{2}': ~u=0,~ ~w'_{0}-\pi /2 \leq v\leq  0
	\label{7.3}
\end{equation}
  \begin{equation}
 B_{3}: ~v-\pi/2 =0,  ~w'_{0}-\pi /2 \leq u\leq  w''_{0}-\pi /2;~~
 B_{3}': ~u=0,~ ~w'_{0}-\pi /2 \leq u\leq  0
	\label{7.4}
\end{equation}

To calculate the phase factor let us give expressions for $f^{\mu}$
on the boundary.
In Minkowski space we have  $f^{\mu}=0$. In all regions
$f^{x}=f^{y}=0$, moreover
 \begin{equation}
f^{u}|_{II}=0,  ~~~f^{v}|_{III}=0
                                                   \label{7.5}
\end{equation}
So the value of the action (\ref{3.11})
for the two plane waves solution (\ref{43.23}), (\ref{43.24})
is reduced to the sum of two terms each of which represents a contribution
from the Cauchy surface $\Sigma '$ and $\Sigma ''$, respectively,
\begin{equation}
S(g_{cl}^{(2pw)})=S^{\Sigma '}+S^{\Sigma ''},
                                                     \label{7.6}
\end{equation}
where
\begin{equation}
S^{\Sigma '}= S'_{2}+S'_{3},~~S^{\Sigma ''}= S''_{2}+S''_{3}+ S''_{4},
	\label{7.7}
\end{equation}
see fig.7. So
\begin {equation} 
                                                          \label {7.71}
<2pw|2pw, WH>={\cal Z}\exp \{-\frac{i}{16\pi G \hbar }(S'_{2}+S'_{3}
-S''_{2}-S''_{3}- S''_{4})\}
\end   {equation} 
here ${\cal Z}$ is the one-loop contribution.

Taking into account the special
form of the metric one can write $f^{w}$
in  regions {\bf II}, {\bf III} and {\bf IV} respectively in the
following form
\begin{equation}
\sqrt {-g}f^{w}|_{II}=\Omega (u)\partial _{u}\ln
[X(u)\Omega ^{2}(u)]
                                            \label{7.8}
\end{equation}
 \begin{equation}
\sqrt {-g}f^{w}|_{III}=\Omega (v)\partial _{v}\ln
[X(v)\Omega ^{2}(v)]                           \label{7.9}
\end{equation}
\begin{equation}
\sqrt {-g}f^{w}|_{IV}=\Omega (w,z)\partial _{w}\ln
[X(w,z)\Omega ^{2}(w,z)]                           \label{7.10}
\end{equation}
$$w=u+v, ~~z=v-u.$$
The representation (\ref {7.8}) was obtained in the following way.
Metric (\ref {43.17})
in $(w,z,x,y)$  coordinates admits the representation
\begin{equation}
g_{\mu\nu}=\left (
\begin{array}{cc}
        g_{\alpha \beta} ~& 0  \\
        0 ~& g_{ij}
\end{array}
\right ).
        \label{7.101}
\end{equation}
with the diagonal matrix $g_{\alpha \beta}$,
\begin{equation}
g_{\alpha \beta}=\left (
\begin{array}{cc}
        g_{ww} ~& 0  \\
        0 ~&g_{zz}
\end{array}
\right ). ,   ~~g_{ww}=-g_{zz}=X
        \label{7.102}
\end{equation}
and  $g_{ij}$ of the form
\begin{equation}
g_{ij}=\Omega \tilde {g_{ij}}                           \label{7.103}
\end{equation}
where
\begin{equation}
                                                        \label{7.104}
  \tilde {g_{ij}}
= \left (
\begin{array}{cc}
     \frac{1}{\chi}  & -\frac{\lambda}{\chi}  \\
        \frac{\lambda}{\chi}& \chi +\frac{\lambda ^{2}}{\chi}
\end{array}
\right ).
\end{equation}
being an element of the group $SL(2)$.
{}From
equation  (\ref {3.3}) it follows
\begin {equation} 
                                                          \label {7.105}
f^{w}=g^{ww}(g^{zz}\partial _{w} g_{ww}+g^{ij}\partial _{w} g_{ij})
\end   {equation} 
Taking into account that the trace  $tr\tilde
{g}^{-1}\partial\tilde {g}=0$ for the group $SL(2)$, we get
\begin {equation} 
                                                          \label {7.106}
f^{w}=g^{ww}(g^{zz}\partial _{w} g_{ww}+2\Omega ^{-1}\partial _{w} \Omega ),
\end   {equation} 
from which follows the representation (\ref {7.10}).

Simple calculations give that
\begin{equation}
S_{2}'=-\frac{1}{16\pi G}\int^{L/2}_{-L/2} dx\int^{L/2}_{-L/2} dy
\int ^{0}_{\pi/2}\int ^{w'_{0}}_{w'_{0}-\pi /2}
dv \delta (u+v -w_{0}')\Omega (u)\partial _{u}\ln
[X(u)\Omega ^{2}(u)]
=
	\label{7.11}
\end{equation}
$$-\frac{1}{16\pi G}L^{2}\int ^{0}_{\pi/2}du\cos ^{2}u\partial _{u}[\ln
((1-p\sin u )^{2}+
q^{2}\sin ^{2}u)\cos ^{4}u]=
$$
$$-\frac{1}{16\pi G}2L^{2}(\int ^{0}_{1}du
\frac{(1-u^{2})(u-p)}{1-2up +u^{2}}-4\int ^{0}_{1}udu)=$$
$$-\frac{1}{16\pi G}(-3-2p+2q^{2}\ln 2(1-p)+4pq \arctan \frac{q}{1-p}) $$
where $L$ is the cut-off in the $x-y$-plane,
\begin{equation}
S_{2}''=-S'_{2}+\bar {S}''_{2},        \label{7.12}
\end{equation}
where
\begin {equation} 
                                                          \label {7.121}
\bar {S}''_{2}=
\frac{L^{2}}{16\pi G}
\int ^{0}_{w''_{0}}du\cos ^{2}u\partial _{u}[\ln ((1-p\sin u )^{2}+
q^{2}\sin ^{2}u)\cos ^{4}u]=
\end   {equation} 
$$\frac{1}{16\pi G}[-3\sin ^{2}w''_{0}-2p\sin w''_{0}+2q^{2}\ln (
\sin ^{2}w''_{0}-2p\sin w''_{0}+1)+4pq \arctan \frac{q\sin w''_{0}}
{1-p\sin w''_{0}}] $$

The part of the boundary which cross the white hole region gives the
following contribution
\begin{equation}
S''_{4}=-\frac{1}{16\pi G}\int^{L/2}_{-L/2} dx\int^{L/2}_{-L/2} dy
\int ^{w'_{0}}_{-w'_{0}}dz\Omega (w,z)\partial _{w}\ln
[X(w,z)\Omega ^{2}(w,z)] =
	\label{7.13}
\end{equation}
$$-\frac{L^{2}}{16\pi G}\int ^{-w''_{0}}_{w''_{0}}dz
[\frac{-2p(1-p\sin w''_{0})\cos ^{2}w''_{0}
\cos z}{(1-p\sin w''_{0})^{2}+q^{2}\sin ^{2}z
}-2\sin  w''_{0}\cos z ] $$
In particular, for  $ w''_{0}=\pi/2$  only the second term in the RHS of
(\ref {7.13}) survives and  we get
$$S^{\Sigma ''}= S''_{4}=\frac{L^{2}}{4\pi G}.$$
Therefore the amplitude describing the transition from two plane waves to
wormhole in the semiclassical approximation is given
\begin {equation} 
                                                          \label {7.15}
<WH|2pw>={\cal Z}\exp \{\frac{i}{ \hbar }(S'_{2}+S'_{3}
+ S''_{4})\}=
\end   {equation} 
$${\cal Z}\exp \{-\frac{iL^{2}}{8\pi G \hbar }[
-1-2p+2q^{2}\ln 2(1-p)+4pq \arctan \frac{q}{1-p}]\}$$

To get the semiclassical answer for the transition amplitude
$<2pw|2pw>$
we have to consider the classical solution
which describes the metric  prolongated as shown in Fig.6.
We see from equations (\ref {7.11}) that $S'_{2}$ does not depend on
$w_{0}'$ and therefore for the transition amplitude describing the elastic
scattering of two plane waves the phase factors  is zero, and
\begin {equation} 
                                                          \label {7.16}
<2pw|2pw>={\cal Z}
\end   {equation} 

$$~$$
{\bf CONCLUSION}
$$~$$
We have considered a possible mechanism for black hole creation in the
 collision  of two light particles
at planckian energies. Many questions deserves a further study.
In particular there is an important question how to relate the momenta
of colliding particles with characteristics of colliding plane waves.
It is known that a plane wave doesn't lead to polarization of vacuum
\cite {Gibb}. It was shown that a plane wave is an exact string background
\cite {AK,NWi,THo,Tse} and there is a duality with extreme black hole
\cite {Rena}. One can conjecture that a dilaton gravity analogue
of the CFX duality between colliding plane waves and non-extremal black holes
discussed in this paper could lead to a corresponding string duality.
$$~$$
{\bf ACKNOWLEDGMENT}
$$~$$
This work has been supported in part by
an operating grant from the National Sciences
and Engineering Research Council of Canada.
I.A. and I.V. thank the Department of Physics for kind hospitality
during their stay at Simon Fraser University.
I.A. and I.V. are
supported in part also by
International Science Foundation under the grant M1L000.
I.V. is grateful to  D.Amati, A.Barvinsky, W.Israel,
V.Frolov, D.Page, G.Veneziano and  A.Zelnikov for stimulating
and critical discussions.
$$~$$

\newpage
\appendix
\section {APPENDIX}
\setcounter{equation}{0}

{\bf Geodesics for colliding plane gravitational waves\\ with zero impact
parameter}

Investigation of geodesics in the metric describing two colliding plane
waves has been considerad in Ref. \cite {}. As has been mentioned above
we are interested in behavior of the null geodesics starting from the surface
$\Sigma '$ (fig.7). Here we prewsent the explicite form of geodesics
in the metric (\ref {6.1})describing two colliding plane waves for zero
impact parameter.  The case of metric (\ref {43.17}) is more complicate
but there is a raison to believe that the behavior of geodesics for the case of
zero impact paremeter reproduces the essential feature of the general case.

Suppose that a geodesic starts in the Minkowski region {\bf I} with
initial momenta $p_{\mu}$.  We investigate its trajectory upto the region
${\bf I}^{'}$ and we express the final 4-momentum in terms of its initial
values in region {\bf I} $(p_{x}, p_{y}$,
$p_{u}, p_{v})$. The geodesic  starting in the region I
at the point $(u^{I}_{0},$ $v^{I}_{0}$, $x^{I}_{0}$, $y^{I}_{0})$
crosses the boundary between the regions I and II
at the point
$(0,$ $v^{II}_{0}$, $x^{II}_{0}$, $y^{II}_{0})$
determined by the following
equations
\begin{eqnarray}
 v^{II}_{0}-v^{I}_{0}    =-\frac{p_{u}}{p_{v}}u^{I}_{0}
 	\label{6.8} \\
 x^{II}_{0}-x^{I}_{0}	 =\frac{2p_{x}m^{2}}{p_{v}}u^{I}_{0}
 \label{6.9}\\
 y^{II}_{0}-y^{I}_{0}	 =\frac{2p_{y}m^{2}}{p_{v}}u^{I}_{0}
 \label{6.10}
\end{eqnarray}

It is well known \cite{Geod} that to get the explicit form of geodesics
\begin{equation}
	\ddot{x}^{\mu}+\Gamma ^{\mu}_{\nu \rho }\dot{x}^{\nu}\dot{x}^{\rho}
	=0,
	\label{6.11}
\end{equation}
it is convenient to use the first order equations
\begin{equation}
m_{0}\dot{x}^{\mu}=g^{\mu\nu}p_{\nu},
	\label{6.12}
\end{equation}
where  $p_{\nu}=\frac{\partial S}{\partial x^{\mu}}$
and the action satisfies the
Hamilton-Jacobi equation
\begin{equation}
g^{\mu\nu}\frac{\partial S}{\partial x^{\mu}}
\frac{\partial S}{\partial x^{\nu}}	=m_{0}^{2}
	\label{6.13}
\end{equation}
 For the metric (\ref {6.1})
in  region
{\bf II} three components of $p_{\mu}$, namely,
$p_{v},p_{x},p_{y},$
are integrals of motions, and
 the Hamilton-Jacobi action $S$ in this region has the following form
in plane wave coordinates
\begin{equation}
S^{II}=\int ^{u}_{u_{0}}du p_{u}	+p_{v}v+p_{x}x+p_{y}y
	\label{6.14}
\end{equation}
where u-component of momentum is given by
\begin{equation}
p_{u}=\frac{m^{2}}{p_{v}}\{ \frac{(1+\sin
u)^{4}}{\cos ^{2}u}p_{x}^{2}+
\frac{1 }{\cos ^{2}u}p_{y}^{2} +(1+\sin
u)^{2}m_{0} ^{2} \}	,
	\label{6.15}
\end{equation}
(mass-shell condition).

Equation (\ref{6.12}) in can be solved explicitly,
\begin{equation}
         x-x^{II}_{0}=-\frac{2m^{2}p_{x}}{p_{v}}\int ^{u}_{u^{II}_{0}}du
	  \frac{(1+\sin
u)^{4}}{\cos ^{2}u}=-\frac{2m^{2}p_{x}}{p_{v}}[8(\tan u -\tan u^{II}_{0})+
	\label{6.16}
	\end{equation}
$$
8(\frac{1}{\cos u}-\frac{1}{\cos u^{II}}_{0})+
4(\cos u -\cos u^{II}_{0})+
\frac{1}{4}(\sin 2u -\cos 2u^{II}_{0}) -\frac{15}{2}(u-u^{II}_{0})],
$$
\begin{equation}
 y-y^{II}_{0}=-\frac{2m^{2}p_{y}}{p_{v}}\int ^{u}_{u^{II}_{0}}
          \frac{du}{\cos ^{2}u}=
          -\frac{2m^{2}p_{y}}{p_{v}}[\tan u -\tan u^{II}_{0}]
	\label{6.18}
\end{equation}
\begin{equation}
 v-v^{II}_{0}=-\frac{p_{x}}{p_{v}}( x-x^{II}_{0})
 -\frac{p_{y}}{p_{v}}y-y^{II}_{0}+
	\label{6.19}
\end{equation}
$$
 m_{0}^{2}[\frac{3}{2}(u-u^{II}_{0})-\frac{1}{2}\cos u (1+\sin u)
 +\frac{1}{2}\cos u^{II}_{0}(1+\sin u^{II}_{0})],
$$
Here the initial data $u^{II} _{0}, v^{II}
_{0},x^{II}_{0},y^{II}_{0}$,
$p_{u,0}^{II},p_{v},p_{x},p_{y}$ for the region {\bf II} are given by
the final data in the region {\bf I} at the boundary between regions
{\bf I} and {\bf II}, and therefore, $u_{0}^{II}=0$.

 The  action $S$ in  regions {\bf IV} and ${\bf IV}^{'}$
in the Schwarzschild coordinates has the following  well-known form
\begin{equation}
S=-Et+l\phi + \int \sqrt{L^{2}-\frac{l^{2}}{\sin ^{2}\theta}}
d\theta +\int
\sqrt{E^{2}-(1-\frac{2m}{r})(m^{2}_{0}+\frac{L^{2}}{r^{2}})}
\frac{dr}{1-\frac{2m}{r}},
	\label{6.20}
\end{equation}
where $l,L$ and $E$ are three integrals of motions  related to the
final data for the region {\bf II} by
\begin{eqnarray}
	  E & =&p_{x}
	\label{6.21} \\
	  l & =&mp_{y}
	\label{6.22} \\
        L^{2} & =&\frac{m^{2}p_{y}^{2}}
        {\cos ^{2} u_{0}}+\{\frac{ m^{2}}{p_{v}}
	[\frac{(1+\sin
u_{0})^{4}}{\cos ^{2}u_{0}}p_{x}^{2}+
\frac{p_{y}^{2}}{\cos ^{2}u_{0}}+m^{2}_{0} (1+\sin
u_{0})^{2}]-p_{v}\}^{2}.
	\label{6.23}
\end{eqnarray}
Here $u_{0}$ determines the point $(u_{0},0)$ in the $(u,v)$ plane at
which the trajectory crosses the boundary between
{\bf II} and {\bf IV}. Let us discuss for simplicity the case $p_{x}=0$.
In this case for null-geodesics the coordinates of the point
$u^{IV}_{0},0,x^{IV}_{0},y^{IV}_{0}$, $u^{IV}_{0}\equiv u_{0}$ are given
in terms of the initial data in the Minkowski region by
\begin{eqnarray}
	u_{0} & = & \arctan (-\frac{p_{v}^{2}}{2mp_{y}^{2}}v^{I}_{0}+
	\frac{1}{2m}u^{I}_{0})
	\label{6.24} \\
	x^{IV}_{0}& = & x^{I}_{0}
	\label{6.25} \\
		y^{IV}_{0}& = & y^{I}_{0}+\frac{v^{I}_{0}}{p_{y}^{2}}
		+\frac{m^{2}p_{y}}{p_{v}}u^{I}_{0}.
	\label{6.26}
\end{eqnarray}

Equations for geodesics in the regions  {\bf IV} and ${\bf IV}^{'}$
for $E=0$ are
\begin{eqnarray}
	 \dot{t}& = & 0
	\label{6.27} \\
	m_{0}\dot{r} & =& -\frac{L}{r}\sqrt{\frac{2m}{r}-1}
	\label{6.28} \\
		m_{0}\dot{\phi} & = & -\frac{1}{r^{2}\sin ^{2} \theta }l
	\label{6.29} \\
		m_{0}\dot{\theta} & =& -\frac{1}{r^{2}}
		\sqrt{L^{2}-l^{2}/\sin ^{2}\theta}
	\label{6.30}
\end{eqnarray}
The trajectory in the $(r,\theta)$ plane is determined by equations
(\ref{6.28})  and (\ref{6.30}) with the initial data
 \begin{equation}
 r|_{\theta = \theta _{0}}=r_{0}, ~~ r_{0}=m(1+\sin u _{0})	.
        \label{6.30'}
 \end{equation}
Integrating,  one finds
\begin{equation}
\arcsin \frac{\cos \theta }{\sqrt{1-l^{2}/L^{2}}}
-\arcsin \frac{\cos \theta _{0}}{\sqrt{1-l^{2}/L^{2}}}=
\arcsin (1-r/m)	-\arcsin (1-r_{0}/m)
	\label{6.31}
\end{equation}
Using the above solution we can find the point $(\pi /2,v_{0})$ in
the $(u,v)$ plane where the  trajectory crosses the boundary between
regions ${\bf IV}^{'}$ and ${\bf II}^{'}$.
Indeed, on the boundary  between
regions ${\bf IV}^{'}$ and ${\bf II}^{'}$ we have
\begin{equation}
\theta '_{0} \equiv \theta |_{u=\pi/2,~v=v_{0}} =\pi -v_{0}
	\label{6.32}
\end{equation}
and
\begin{equation}
r'_{0} \equiv r|_{u=\pi/2,~ v=v_{0}} =m(1+\cos v_{0}).
	\label{6.33}
\end{equation}
Equation (\ref {6.31}) also gives the relation between $r'_{0} $
and $\theta '_{0} $. Substituting in
that relation the expressions of $r_{0} $
and $r'_{0}$  in term of $u_{0}$ and $v_{0}$ ( equations (\ref {6.30
}) and (\ref {6.33}) respectively) and $\theta _{0}=\pi /2+u_{0}$ and
(\ref {6.32})  we get
\begin{equation}
-       \frac{\cos v_{0}}{\sqrt{1-
l^{2}/L^{2}}}\sqrt{1-\frac{\sin ^{2}u_{0}}
{1- l^{2}/L^{2}}}+
        \frac{\sin u_{0}}{\sqrt{1-
        l^{2}/L^{2}}}\sqrt{1-\frac{\cos ^{2}v_{0}}
{1- l^{2}/L^{2}}}= -\cos (v_{0}+u_{0})
	\label{6.34}
\end{equation}

Let us look at the geodesics in the $(\tau ,\zeta)$
plane of the Kruskal coordinates. From equations (\ref {6.27}-\ref {6.30})
and (\ref {6.6}) it follows that
\begin{equation}
\frac{d\zeta}{d\tau}=\frac{\zeta}{\tau}	,
	\label{6.35}
\end{equation}
which yields
\begin{equation}
\zeta \tau=\frac{\zeta _{0}}{\tau _{0}} \tau	,
	\label{6.36}
\end{equation}
where
\begin{equation}
\zeta _{0}=- a(r(t_{0}))\sinh t_{0}/4m,
        \label{6.36'}
\end{equation}
Taking a suitable initial data in the Minkowski region {\bf I} we can set
$\zeta _{0}=0$ (this choice corresponds to motion in the $x=0$ plane).
All geodesics then pass through the wormhole in the Kruskal diagram.

In region ${\bf II}^{'}$ the geodesics are similar to those in the
 region {\bf II}  with $p_{v}, p_{u}$ interchanged, i.e.
 \begin{equation}
S^{II'}=\int ^{u}_{u_{0}}du \tilde{p}_{u}
+\tilde{p}_{v}v+p_{x}x+p_{y}y
	\label{6.37}
\end{equation}
where the $v$-component of momentum is given by
\begin{equation}
\tilde{p}_{v}=\frac{m^{2}}{\tilde{p}_{u}}\{ \frac{(1+\cos
v)^{4}}{\sin ^{2}v}p_{x}^{2}+
\frac{(1 }{\sin ^{2}v}p_{y}^{2} +(1+\cos
v)^{2}m_{0} ^{2} \}	,
	\label{6.38}
\end{equation}
$\tilde{p}_{u}$ can be found from the equation
\begin{equation}
\frac{4m^{4}p^{4}_{y}}{p_{v}^{2}\cos ^{4}u_{0}}-
\frac{4m^{4}p^{4}_{y}}{\tilde{p}_{u}^{2}\sin  ^{4}v_{0}}+p^{2}_{v}
-\tilde{p}_{u}^{2}+ 2m^{2}
(\frac{p^{2}_{y}}{\cos ^{2}u_{0}}-
\frac{p^{2}_{y}}{\sin  ^{4}v_{0}})
	\label{6.39}
\end{equation}
which follows from junction conditions on
the boundary between  {\bf II}
and
{\bf IV} and  ${\bf II}^{'}$
and
{\bf IV} regions.  $u_{0}$ and $v_{0}$ in (\ref {6.39}) are defined by
equations (\ref {6.24}) and (\ref {6.34}).

{}From above consideration follows that all null
geodesics starting from the Cauchy surface $\Sigma '$ reach the surface
$\Sigma ''$ or end up on the null surfaces $u=\pi /2, v<0; $  and
$v=\pi /2, u<0 $.

\newpage

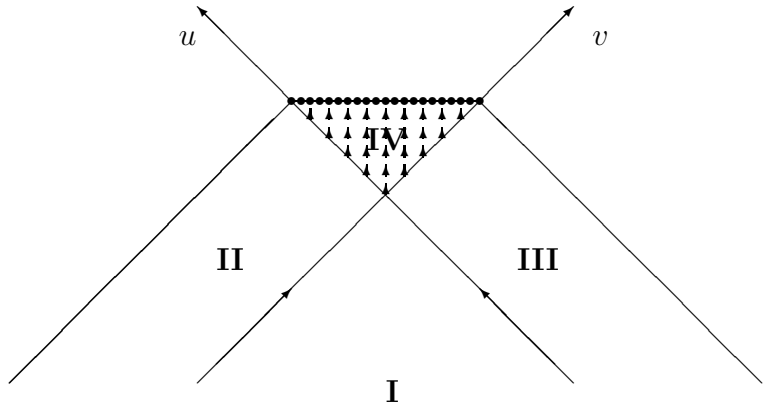
\begin{figure}
\setlength{\unitlength}{0.5cm}
\begin{center}
\begin{picture}(15,15)(-5,-15)
  \put(-2.5,2.5){\line(1,0){5.0}}
  \multiput(-2.5,2.5)(0.25,0.0){21}{\circle*{0.2}}
\put(-5.0,-5.0){\vector(1,1){10.0} }
\put(-4.5,-4.5){\vector(1,1){2.0} }
\put(4.5,-4.5){\vector(-1,1){2.0} }
\put(5.0,-5.0){\vector(-1,1){10.0} }
        \put(-10.0,-5.0){\line(1,1){7.5}}
 \put(-10.0,-5.0){\line(1,1){7.0}}
\put(10.0,-5.0){\line(-1,1){7.5}}
\multiput(0.0,0.0)(0.0,0.5){5}{\vector(0,1){0.3}}
\multiput(0.5,0.5)(0.0,0.5){4}{\vector(0,1){0.3}}
\multiput(1.0,1.0)(0.0,0.5){3}{\vector(0,1){0.3}}
\multiput(1.5,1.5)(0.0,0.5){2}{\vector(0,1){0.3}}
\multiput(2.0,2.0)(0.0,0.5){1}{\vector(0,1){0.3}}
\multiput(-0.5,0.5)(0.0,0.5){4}{\vector(0,1){0.3}}
\multiput(-1.0,1.0)(0.0,0.5){3}{\vector(0,1){0.3}}
\multiput(-1.5,1.5)(0.0,0.5){2}{\vector(0,1){0.3}}
\multiput(-2.0,2.0)(0.0,0.5){1}{\vector(0,1){0.3}}

 \put(5.5,4.0){\mbox {$v$}}
 \put(-5.5,4.0){\mbox {$u$}}
\put(-4.5,-2.0){\mbox {{\bf II}}}
\put(3.5,-2.0){\mbox {{\bf III}}}
\put(0.0,-5.5){\mbox {{\bf I}}}
\put(-8,-7.0){\mbox {$(ds^{I})^{2}=4m^{2}dudv-dx^{2}-dy^{2},
	 $}}
	 \put(-12,-9.0){\mbox {$(ds^{II})^{2}=
 4m^{2}[1+\sin u]  dudv -
 \cos ^{2}u [1+\sin u]^{-2}dx^{2}
 -\cos ^{2}u [1+\sin u]^{2}dy^{2}
	 $}}
	 \put(-12,-11.0){\mbox {$(ds^{III})^{2}=
 4m^{2}[1+\sin v]  dudv -
 \cos ^{2}v [1+\sin v]^{-2}dx^{2}
 -\cos ^{2}v [1+\sin v]^{2}dy^{2}
	 $}}
	 \put(-12,-13.0){\mbox {$(ds^{IV})^{2}=
 4m^{2}[1+\sin (u+v)]  dudv -
 \cos ^{2}(u+v) [1+\sin (u+v)]^{-2}dx^{2}
	 $}}
	 \put(-8,-15.0){\mbox {$ -\cos ^{2}(u-v) [1+\sin (u+v)]^{2}dy^{2}
 $}}
\put(-0.5,1.2){\mbox {{\bf IV}}}
\end{picture}
 \end{center}
\caption{$(u,v)$ plane wave coordinates }\label{f3}
\end{figure}

\begin{figure}
\setlength{\unitlength}{0.5cm}
\begin{center}
\begin{picture}(10,10)(-5,-5)
  \put(-2.5,2.5){\line(1,0){5.0}}
  \multiput(-2.5,2.5)(0.5,0.0){11}{\circle*{0.2}}
  \multiput(-0.5,2.0)(0.5,0.0){11}{\circle*{0.2}}
  \multiput(1.5,1.5)(0.5,0.0){11}{\circle*{0.2}}
  \multiput(3.5,1.0)(0.5,0.0){11}{\circle*{0.2}}
  \multiput(5.5,0.5)(0.5,0.0){11}{\circle*{0.2}}
  \multiput(7.5,0.0)(0.5,0.0){11}{\circle*{0.2}}
  \multiput(9.5,-0.5)(0.5,0.0){11}{\circle*{0.2}}
\put(-5.0,-5.0){\vector(1,1){10.0} }
\put(5.0,-5.0){\vector(-1,1){10.0} }
\put(0.0,0.0){\vector(4,-1){14.0} }
\put(-2.5,2.5){\line(4,-1){12.0}}
 \put(2.5,2.5){\line(4,-1){12.0}}
  \put(12.0,-3.0){\line(1,1){2.5}}
   \put(11.5,-2.5){\line(1,1){2.0}}
   \put(11.0,-2.0){\line(1,1){1.5}}
    \put(10.5,-1.5){\line(1,1){1.0}}
     \put(10.0,-1.0){\line(1,1){0.5}}
   \put(12.0,-3.0){\line(-1,1){2.5}}
   \put(9.5,-0.5){\line(1,0){5.0}}
  \put(-10.0,-5.0){\line(1,1){7.5}}
\put(10.0,-5.0){\line(-1,1){7.5}}
 \put(14.0,-4.0){\mbox {$x$}}
 \put(5.5,4.0){\mbox {$v$}}

 \put(-5.5,4.0){\mbox {$u$}}
\put(-4.5,-2.0){\mbox {{\bf II}}}
\put(3.5,-2.0){\mbox {{\bf III}}}
\put(0.0,-5.5){\mbox {{\bf I}}}
\put(-0.5,1.2){\mbox {{\bf IV}}}
\end{picture}
 \end{center}
\caption{$(x,u,v)$ plane wave coordinates }\label{f4}
\end{figure}
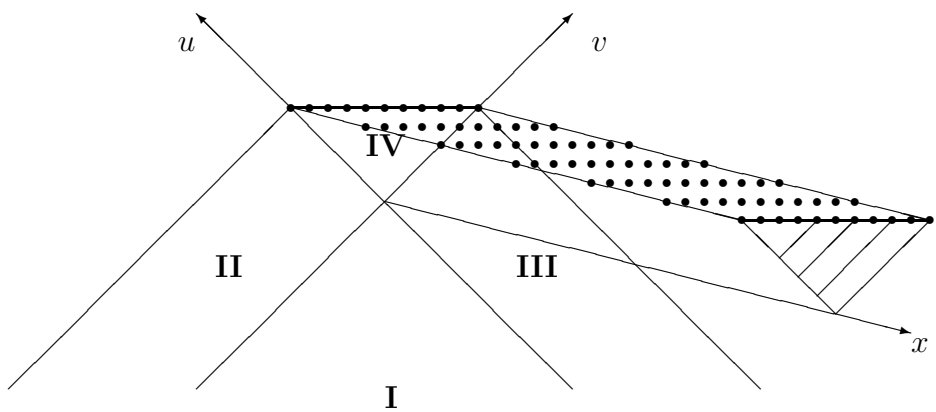

\newpage

\begin{figure}
\setlength{\unitlength}{0.5cm}
\begin{center}
\begin{picture}(10,20)(-5,-18)
  \put(-7.5,-7.5){\line(1,1){15.0}}
 \put(7.5,-7.5){\line(-1,1){15.0}}
 \put(-7.5,0.0){\vector(1,0){15.0} }
\put(0.0,-7.5){\vector(0,1){15.0} }
\put(0.1,7.5){\mbox {$\tau$}}
\put(7.5,7.0){\mbox {$r=2m$}}
\put(6.8,7.5){\mbox {$r=m$}}
\put(6.5,8.0){\mbox {$r=0$}}
\put(7,-0.5){\mbox {$\zeta$}}
\multiput(-6.5,7.5)(0.1,-0.1){12}{\circle*{0.1}}
\multiput(-5.5,6.5)(0.1,-0.075){21}{\circle*{0.1}}
\multiput(-3.5,5.0)(0.1,-0.05){10}{\circle*{0.1}}
\multiput(-2.5,4.5)(0.1,-0.025){14}{\circle*{0.1}}
\multiput(-1.1,4.15)(0.1,-0.01){11}{\circle*{0.1}}
\multiput(6.5,7.5)(-0.1,-0.1){12}{\circle*{0.1}}
\multiput(5.5,6.5)(-0.1,-0.075){21}{\circle*{0.1}}
\multiput(3.5,5.0)(-0.1,-0.05){10}{\circle*{0.1}}
\multiput(2.5,4.5)(-0.1,-0.025){14}{\circle*{0.1}}
\multiput(1.1,4.15)(-0.1,-0.01){11}{\circle*{0.1}}

\multiput(6.5,-7.5)(-0.1,0.1){12}{\circle*{0.1}}
\multiput(5.5,-6.5)(-0.1,0.075){21}{\circle*{0.1}}
\multiput(3.5,-5.0)(-0.1,0.05){10}{\circle*{0.1}}
\multiput(2.5,-4.5)(-0.1,0.025){14}{\circle*{0.1}}
\multiput(1.1,-4.15)(-0.1,0.01){11}{\circle*{0.1}}

\multiput(-6.5,-7.5)(0.1,0.1){12}{\circle*{0.1}}
\multiput(-5.5,-6.5)(0.1,0.075){21}{\circle*{0.1}}
\multiput(-3.5,-5.0)(0.1,0.05){10}{\circle*{0.1}}
\multiput(-2.5,-4.5)(0.1,0.025){14}{\circle*{0.1}}
\multiput(-1.1,-4.15)(0.1,0.01){11}{\circle*{0.1}}

\multiput(-6.5,7.0)(0.2,-0.2){6}{\circle{0.05}}
\multiput(-5.5,6.0)(0.2,-0.15){11}{\circle{0.05}}
\multiput(-3.5,4.5)(0.2,-0.1){6}{\circle{0.05}}
\multiput(-2.5,4.0)(0.2,-0.07){7}{\circle{0.05}}
\multiput(-1.1,3.55)(0.2,-0.04){6}{\circle{0.05}}
\multiput(6.5,7.0)(-0.2,-0.2){6}{\circle{0.05}}

\multiput(5.5,6.0)(-0.2,-0.15){11}{\circle{0.05}}
\multiput(3.5,4.5)(-0.2,-0.1){6}{\circle{0.05}}
\multiput(2.5,4.0)(-0.2,-0.07){7}{\circle{0.05}}
\multiput(1.1,3.55)(-0.2,-0.04){6}{\circle{0.05}}
\multiput(6.5,-7.0)(-0.2,0.2){6}{\circle{0.05}}
\multiput(5.5,-6.0)(-0.2,0.15){11}{\circle{0.05}}
\multiput(3.5,-4.5)(-0.2,0.1){6}{\circle{0.05}}
\multiput(2.5,-4.0)(-0.2,0.07){7}{\circle{0.05}}
\multiput(1.1,-3.55)(-0.2,0.04){6}{\circle{0.05}}
\put(3.5,-4.5){\line(1,1){0.5}}
\put(2.5,-4.0){\line(1,1){0.5}}
\put(2.2,-3.8){\line(1,1){0.5}}
\put(1.1,-3.55){\line(1,1){1.0}}
\put(0.5,-3.5){\line(1,1){1.0}}


\multiput(-6.5,-7.0)(0.2,0.2){6}{\circle{0.05}}
\multiput(-5.5,-6.0)(0.2,0.15){11}{\circle{0.05}}
\multiput(-3.5,-4.5)(0.2,0.1){6}{\circle{0.05}}
\multiput(-2.5,-4.0)(0.2,0.07){7}{\circle{0.05}}
\multiput(-1.1,-3.55)(0.2,0.04){6}{\circle{0.05}}
\put(-3.5,-4.5){\line(1,1){3.8}}
\put(-2.9,-3.5){\line(1,1){2.8}}
\put(-2.5,-4.0){\line(1,1){3.0}}
\put(-1.1,-3.55){\line(1,1){2.2}}
\put(0.0,-3.4){\line(1,1){1.5}}

$\multiput(-2.2,-3.8)(0.2,0.2){7}{\line(1,1){0.1}}
$\multiput(-1.1,-3.55)(0.2,0.2){5}{\line(1,1){0.1}}
\put(1.5,-3.2){\mbox {{\bf IV}}}
\put(1.5,2.7){\mbox {{\bf IV'}}}
\multiput(0.0,-3.4)(0.0,0.5){6}{\vector(0,1){0.3}}
	 \put(-6,-13.0){\mbox {$ds^{2}=\frac{32m^{3}}{r}e^{-r/2m}
	 (d\tau ^{2} -d\zeta ^{2})
	 -r^{2}(d\theta ^{2} +\sin ^{2} \theta d\phi ^{2})
	 $}}
	 \put(-4,-15.0){\mbox {$\tau ^{2} -\zeta ^{2} =a^{2}(r),~~a(r)=(1-r/2m)
	 ^{1/2}e^{r/4m}$}}
\end{picture}
 \end{center}
\caption{Kruskal coordinates}\label{f5}
\end{figure}

\newpage

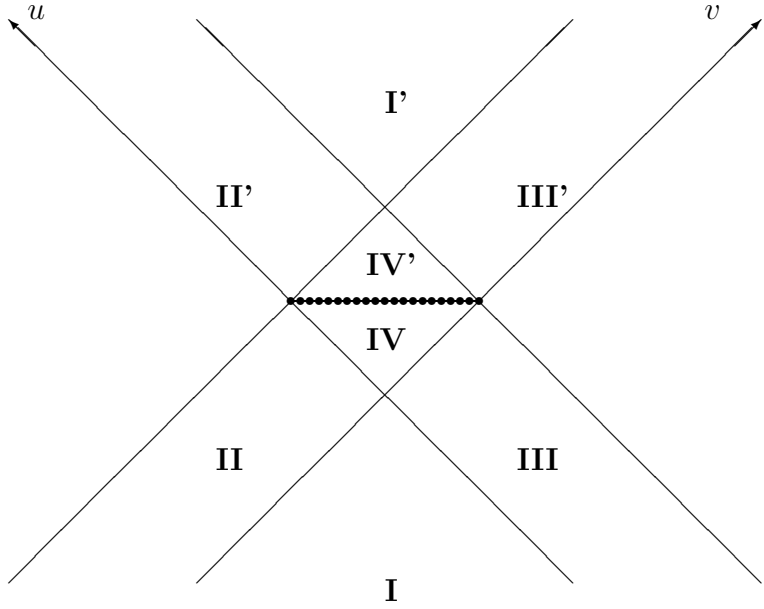
\begin{figure}
\setlength{\unitlength}{0.5cm}
\begin{center}
\begin{picture}(17,20)(-5,-15)
  \put(-2.5,2.5){\line(1,0){5.0}}
  \multiput(-2.5,2.5)(0.25,0.0){21}{\circle*{0.2}}
\put(-5.0,-5.0){\vector(1,1){15.0} }
\put(5.0,-5.0){\vector(-1,1){15.0} }
        \put(-10.0,-5.0){\line(1,1){15.0}}
\put(10.0,-5.0){\line(-1,1){15.0}}
 \put(8.5,10.0){\mbox {$v$}}
 \put(-9.5,10.0){\mbox {$u$}}
\put(-4.5,-2.0){\mbox {{\bf II}}}
\put(3.5,-2.0){\mbox {{\bf III}}}
  \put(3.5,5.0){\mbox {{\bf III'}}}
  \put(-4.5,5.0){\mbox {{\bf II'}}}
\put(0.0,-5.5){\mbox {{\bf I}}}
\put(0.0,7.5){\mbox {{\bf I'}}}
\put(-0.5,1.2){\mbox {{\bf IV}}}
\put(-0.5,3.2){\mbox {{\bf IV'}}}
\put(-8,-7.0){\mbox {$(ds^{I'})^{2}=4m^{2}dudv-dx^{2}-dy^{2},
	 $}}
	 \put(-12,-9.0){\mbox {$(ds^{II'})^{2}=
 4m^{2}[1+\cos v]  dudv -
 \sin ^{2}v [1+\cos v]^{-2}dx^{2}
 -\sin ^{2}v [1+\cos v]^{2}dy^{2}
	 $}}
         \put(-12,-11.0){\mbox {$(ds^{III'})^{2}=
 4m^{2}[1+\sin u]  dudv -
 \sin ^{2}u [1+\cos u]^{-2}dx^{2}
 -\sin ^{2}u [1+\cos u]^{2}dy^{2}
	 $}}
	 \put(-12,-13.0){\mbox {$(ds^{IV'})^{2}=
 4m^{2}[1+\sin (u+v)]  dudv -
 \cos ^{2}(u+v) [1+\sin (u+v)]^{-2}dx^{2}
	 $}}
	 \put(-8,-15.0){\mbox {$ -\cos ^{2}(u-v) [1+\sin (u+v)]^{2}dy^{2}
 $}}
\end{picture}
 \end{center}
\caption{Extended plane wave coordinates}\label{f6}
\end{figure}

\begin{figure}
\setlength{\unitlength}{0.5cm}
\begin{center}
\begin{picture}(8,10)(-5,-5)
  \put(-2.5,2.5){\line(1,0){5.0}}
  \multiput(-2.5,2.5)(0.25,0.0){21}{\circle*{0.2}}
\put(-5.0,-5.0){\vector(1,1){10.0} }
\put(5.0,-5.0){\vector(-1,1){10.0} }
        \put(-10.0,-5.0){\line(1,1){7.5}}
 \put(-10.0,-5.0){\line(1,1){7.0}}
\put(-10.0,-4.0){\line(1,0){20.0}}
\put(-4.5,1.0){\line(1,0){9.0}}
\put(10.0,-5.0){\line(-1,1){7.5}}
 \put(5.5,4.0){\mbox {$v$}}
 \put(-5.5,4.0){\mbox {$u$}}
 \put(-5.5,1.2){\mbox {$\Sigma ''$}}
 \put(4.5,1.0){\mbox {$w_{0} ''$}}
 \put(-0.5,1.2){\mbox {$S_{4}$}}
 \put(2.5,1.2){\mbox {$S''_{3}$}}
 \put(-2.8,1.2){\mbox {$S''_{2}$}}
\put(-10.5,-3.0){\mbox {$\Sigma '$}}
 \put(10.5,-4.0){\mbox {$w_{0}'$}}
\put(6.5,-3.8){\mbox {$S'_{3}$}}
  \put(-6.5,-3.8){\mbox {$S'_{2}$}}
  \put(-8.5,-2.0){\mbox {$B_{2}$}}
 \put(-1.5,-2.0){\mbox {$B'_{2}$}}
  \put(7.5,-2.0){\mbox {$B_{3}$}}
 \put(0.5,-2.0){\mbox {$B'_{3}$}}

\put(-4.5,-2.0){\mbox {{\bf II}}}
\put(3.5,-2.0){\mbox {{\bf III}}}
\put(0.0,-5.5){\mbox {{\bf I}}}
\end{picture}
 \end{center}
\caption{Initial and final Cauchy surfaces}\label{f7}
\end{figure}
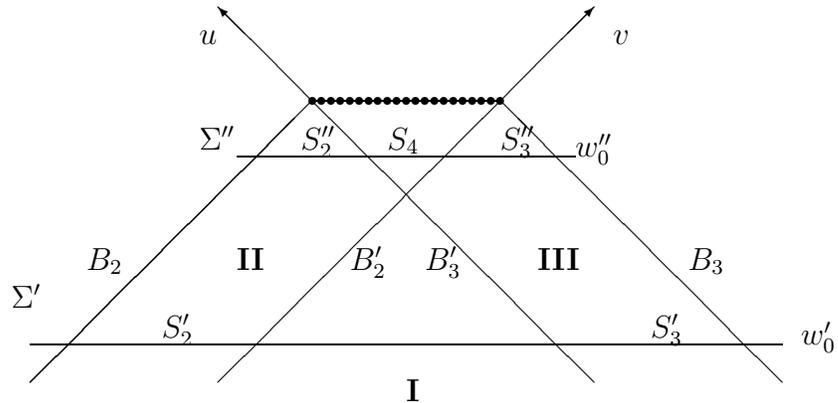
\end{document}